**Silver Staining of 2D Electrophoresis Gels**

Thierry Rabilloud

CEA-DSV-iRTSV/LCBM and UMR CNRS-UJF 5249

CEA Grenoble

17 rue des martyrs, F-38054 Grenoble Cedex 9, France

thierry.rabilloud@cea.fr

Running Head: Silver staining

**i. Abstract**

Silver staining is used to detect proteins after electrophoretic separation on polyacrylamide gels. It combines excellent sensitivity (in the low nanogram range) with the use of very simple and cheap equipment and chemicals. For its use in proteomics, two important additional features must be considered, compatibility with mass spectrometry and quantitative response. Both features are discussed in this chapter, and optimized silver staining protocols are proposed.

## 1. Introduction

Silver staining of polyacrylamide gels was introduced in 1979 by Switzer et al. *(1)*, and rapidly gained popularity owing to its high sensitivity, ca. 100 times higher than staining with classical Coomassie Blue and 10 times higher than colloidal Coomassie Blue. However, the first silver staining protocols were not trouble-free. High backgrounds and silver mirrors were frequently experienced, with a subsequent decrease in sensitivity and reproducibility. This led many authors to suggest improved protocols, so that more than 100 different silver staining protocols for proteins in polyacrylamide gels can be found in the literature. However, all of them are based on the same principle (*see (2)* and *(3)* for details) and comprise four major steps.

a) The first step is fixation, it precipitates the proteins in the gels and removes at the same time the interfering compounds present in the 2D gels (glycine, Tris, SDS and carrier ampholytes present in 2D gels).

b) The second step is sensitization, and aims at increasing the subsequent image formation. Numerous compounds have been proposed for this purpose. All these compounds bind to the proteins, and are also able either to bind silver ion, or to reduce silver ion into metallic silver, or to produce silver sulfide *(2, 3)*.

c) The third step is silver impregnation. Either plain silver nitrate or ammoniacal silver can be used, but nowadays silver nitrate is more extensively used (*see* **Note 1**).

d) The fourth last step is image development. For gels soaked with silver nitrate, the developer contains formaldehyde, carbonate and thiosulfate. The use of the latter compound, introduced by Blum et al. *(4)*, reduces dramatically the background and allows for thorough development of the image. When the desired image level is obtained, development is stopped by dipping the gel in a stop solution, generally containing acetic acid and an amine to reach a pH of 7. Final stabilization of the image is achieved by thorough rinsing in water to remove all the compounds present in the gel.

However, the development of downstream protein characterization methods, such as analysis by mass spectrometry starting from gel-separated proteins, has brought several constraints to the forefront. The first is the interface with mass spectrometry, and this encompasses the compatibility with enzymatic digestion and peptide extraction, as well as the absence of staining-induced peptide modifications. While the exquisite sensitivity of silver staining is unanimously recognized, its compatibility with downstream analysis appears more problematic than staining with organic dyes (e.g. Coomassie Blue). A mechanistic study *(5)* has shown that these problems are linked in part to the pellicle of metallic silver deposited on the proteins during staining, but are mainly due to the presence of formaldehyde during silver staining. Up to now, formaldehyde is the main chemical known able to produce a silver image of good quality in protein staining, and attempts to use other chemicals have proven rather unsuccessful. However, besides a lowered peptide representation in silver stained gels *(5)*, formaldehyde induces a host of peptide modifications *(6, 7)* as well as formylation *(8)*, the latter being most likely caused by the formic acid produced upon reaction of formaldehyde with silver ions in the image development step. In order to minimize these problems, destruction of the remaining formaldehyde by oxidation *(9)* is highly recommended, and this should take place as early as possible after image development and spot excision *(5)*. These problems are common to all silver staining protocols, although their extent is variable from one protocol to another. Some guidelines for the choice of a silver staining protocol are described in **Note 1**. To alleviate the problems linked with the use of formaldehyde, a new protocol using reducing sugars as developing agents has been recently described *(10)*. This protocol is also detailed in this chapter and typical gels stained by the protocols described are shown in **Fig. 1**. As artefacts arising from grafting of the sugars cannot be excluded and be confused with protein glycation occurring in vivo, pentoses such as xylose can be used for silver staining development, as they will give a peptide mass increase different than that induced by in vivo glycation.

The second constraint very prevalent in proteomics is linked to the fact that stained 2D gels are used to perform quantitative analysis. In addition to sensitivity, this puts special emphasis on stain linearity and homogeneity from run to run.

In fact, silver staining is linear over a rather limited dynamic range, as show in **Fig. 2** and **3** on 1D and 2D gels, respectively. **Fig. 2** shows in particular two important features of silver staining, i.e. the rather poor linearity at very low intensities, and the variable slope of the dose-signal curve from one protein to the other. However, it must be emphasized that the slope of the response curve is always lesser or equal to 1. This means in turn that silver staining is a conservative technique that will in most times underestimate the variations in protein abundances.

Moreover, modern silver staining is no longer an erratic technique, as it was in its infancy *(1)*. This is shown on **Fig. 4**, which compares the variability of a series of 2D gels run on the same sample and stained by various methods. It can be shown that the dispersion of the signals (a measure of variability) is not greater with adequate silver staining than with Coomassie Blue.

This variability can be further decreased by the used of batch apparatus for silver staining *(11)*.

## 2. Materials

### 2.1. Equipment

1. Glass dishes or polyethylene food dishes. The latter are less expensive, have a cover and can be easily piled up for multiple staining. They are however more difficult to clean, and it is quite important to avoid scratching of the surface, which will induce automatic silver deposition in subsequent stainings. Traces of silver are generally easily removed by wiping the plastic box with a tissue soaked with ethanol. If this treatment is not sufficient, stains are easily removed with Farmer's reducer (0.1% sodium carbonate, 0.3% potassium hexacyanoferrate (III) and 0.6% sodium thiosulfate). Thorough rinsing of the box with water and ethanol terminate the cleaning process.
2. Plastic sheets (e.g. the thin polycarbonate sheets sold by Bio-Rad for multiple gel casting) used for batch processing.
3. Reciprocal shaking platform: The use of orbital or three-dimensional movement shakers is not recommended.

### 2.2. Reagents

Generally speaking, chemicals are of standard pro analysis grade.

1. Water: The quality of the water is of great importance. Water purified by ion exchange cartridges, with a resistivity greater than 15 MegaOhms/cm, is very adequate, while distilled water gives more erratic results.
2. Formaldehyde: Formaldehyde stands for commercial 37-40% formaldehyde. This is stable for months at room temperature. It should not be stored at 4°C, as this increases polymerization and deposition of formaldehyde. The bottle should be discarded when a layer of polymer is visible at the bottom of the bottle.
3. Sodium thiosulfate solution: 10% (w/v) solution of crystalline sodium thiosulfate pentahydrate in water. Small volumes of this solution (e.g. 10 ml) should be prepared fresh every week and stored at room temperature.
4. Ethanol: A technical grade of alcohol can be used, and 95% ethanol can be used instead of absolute ethanol without any volume correction. The use of denatured alcohol is however not recommended.
5. Silver nitrate solution: 1 *N* silver nitrate. A 1 *N* silver nitrate solution (e.g. from Fluka) is less expensive than solid silver nitrate, and is stable for months if kept in a fridge

(protection from light is necessary).

### 2.2.1. Solutions for PROTOCOL 1 (fast silver staining)

1. Fix solution: 5% acetic acid, 30% ethanol (*see* **Note 2**).
2. Sensitivity enhancing solution: 2 ml of 10% sodium thiosulfate solution per liter.
3. Silver stain solution: 12.5 ml of 1 *N* silver nitrate solution per liter.
4. Development solution I: 30 g anhydrous potassium carbonate, 250 µl of 37% formaldehyde and 125 µl of 10% thiosulfate solution per liter.
5. Stop solution: 40 g of Tris and 20 ml of acetic acid (100%) per liter.

### 2.2.2. Solutions for PROTOCOL 2 (silver staining with aldose-based developer)

1. Fix solution: 5% acetic acid, 30% ethanol (*see* **Note 2**).
2. Sensitivity enhancing solution: 2 ml of 10% sodium thiosulfate solution per liter.
3. Silver stain solution: 12.5 ml of 1 *N* silver nitrate solution per liter.
4. Development solution II: 0.1 *M* boric acid, 0.15 *M* sodium hydroxide, 2% (w/v) galactose or xylose and 125 µl of 10% thiosulfate solution per liter.
5. Stop solution: 40g of Tris and 20 ml of acetic acid per liter.

### 2.2.3. Destaining solution for ADDITIONAL PROTOCOL (spot destaining prior to mass spectrometry)

1. Prepare a 30 m*M* potassium ferricyanide solution in water. Additionally, prepare a 100 m*M* sodium thiosulfate solution.
   CAUTION: both solutions must be prepared the day of use.
2. Just before use, mix equal volumes of the potassium ferricyanide and thiosulfate solutions.
   CAUTION: This resulting yellowish solution is stable and active for less than 30 minutes, and must be used immediately.

## 3. Methods

### 3.1. General practice

Batches of gels (up to 4 gels per box) can be stained. For a batch of 3 or 4 medium-sized gels (e.g. 160x200x1.5 mm), 1 liter of the required solution is used, which corresponds to a solution/gel volume ratio of at least 5. 500 ml of solution is used for 1 or 2 gels. Batch processing can be used for every step longer than 5 minutes, except for image development, where one gel per box is required. For steps shorter than 5 minutes, the gels should be dipped individually in the corresponding reagent(s).

For changing solutions, the best way is to use a plastic sheet. This is pressed on the pile of gels with the aid of a gloved hand. Inclining the entire setup allows to empty the box while keeping the gels in it. The next solution is poured with the plastic sheet in place, which prevents the flow to fold or break the gels. The plastic sheet is removed after the solution change and kept in a separate box filled with water until the next solution change. This water is changed after each complete round of silver staining.

When gels must be handled individually, they are manipulated with gloved hands (*see* **Note 3**). Except for development or short steps, where occasional hand agitation of the staining vessel is convenient, constant agitation is required for all the steps. A reciprocal ("ping-pong") shaker is used at 30-40 strokes per minutes.

Two different silver staining protocols are detailed below. The rationale for choosing one of them according to the constraints brought by the precise 2D protocol used and the requisites of the experimentator are described in **Note 1**.

### 3.2. PROTOCOL 1: fast silver staining

This protocol is based on the protocol of Blum et al *(4)*, with modifications *(12, 13)*.

1. Soak the gels in fix solution for at least 3x 30 minutes or overnight with one solution change for 2D gels (*see* **Note 4**).
2. Rinse in water for 3x 10 minutes.
3. To sensitize, soak gels for 1 minute (1 gel at a time) in sensitivity enhancing solution.
4. Rinse 2x 1 minute in water (*see* **Note 5**).
5. Impregnate for at least 30 minutes in silver solution (*see* **Note 6**).
6. Rinse in water for 5-15 seconds (*see* **Note 7**).
7. Develop image (10-20 minutes) in development solution I (*see* **Note 8** and **9**).
8. Stop development (30-60 minutes) in stop solution.
9. Rinse with water (several changes) prior to drying or densitometry.

### 3.3. PROTOCOL 2: silver staining with aldose-based developer

1. Soak the gels in fix solution for at least 3x 30 minutes or overnight with one solution change for 2D gels (*see* **Note 4**).
2. Rinse in water for 3x 10 minutes.
3. To sensitize, soak gels for 1 minute (1 gel at a time) in sensitivity enhancing solution.
4. Rinse 2x 1 minute in water (*see* **Note 5**).

5. Impregnate for at least 30 minutes in silver solution (*see* **Note 6**).
6. Rinse in water for 5-15 seconds (*see* **Note 7**).
7. Develop image (10-20 minutes) in development solution II (*see* **Notes 8**, **9** and **10**).
8. Stop development (30-60 minutes) in stop solution.
9. Rinse with water (several changes) prior to drying or densitometry.

### 3.4. ADDITIONAL PROTOCOL: spot destaining prior to mass spectrometry

Silver staining interferes strongly with mass spectrometry analysis of spots or bands excised from stained electrophoresis gels. This interference can be reduced by destaining the spots or bands prior to the standard digestion protocols. The destaining protocols giving minimal artifacts are the ferricyanide-thiosulfate protocol of Gharahdaghi et al. *(9)*. This protocol can be carried out on spots or bands in microtubes (0.5 or 1.5 ml) or in 96 well plates. The use of a shaking device (plate shaker or rotating wheel for tubes) is recommended.

**Procedure**

1. Cover the spots or bands with 0.15 ml of spot destaining solution. The stain should be removed in 5-10 minutes.
2. Remove the solution, and rinse the spots 5x5 minutes with water (0.15 ml per gel piece).
3. Remove the water, and soak the gel pieces in 200 m*M* ammonium hydrogenocarbonate (in water) for 20 minutes (0.15 ml per gels piece).
4. Repeat step 2.
5. Process the rinsed gel pieces for mass spectrometry, or store dry at –20°C until use.

### 4. Notes

1. From the rather simple theoretical bases described in the introduction, more than 100 different protocols were derived. The changes from one protocol to another are present either in the duration of the different steps, or in the composition of the solutions. The main variations concern both the concentration of the silver reagent or the nature and concentration of the sensitizers. Only a few comparisons of silver staining protocols have been published *(12)*. From these comparisons, selected protocols have been proposed in the former sections. The choice of a protocol will depend on the constraints of the experimental setup and of the requisites of the experimentator (speed, reproducibility, compatibility with mass spectrometry etc.). Although they can be very sensitive for basic proteins, we have excluded from this selection protocols using silver-ammonia, as the results are fairly dependent on the ammonia/silver concentration. Furthermore, home-made gels with included thiosulfate must be used for optimal results *(14)*, and Tricine-based gels cannot be used. These restrictions have thus driven us to consider in priority protocols using plain silver nitrate or staining, as they are more robust and versatile.

2. Other fixation processes can be used. For gels running overnight, the procedure can be shortened. For silver nitrate staining, fixation can be reduced to a single 30 minutes bath *(15)*. This will improve sequence coverage in mass spectrometry, at the expense of a strong chromatism (spots can be yellow, orange, brown or grey), making image analysis difficult. Furthermore, ampholytes are not removed by short fixation and give a grey background at the bottom of the 2D gels. Thorough removal of ampholytes requires an overnight fixation

3. The use of powder-free, nitrile, gloves is strongly recommended, as standard gloves are often the cause of pressure marks.

4. The fixation process can be altered if needed. The figures indicated in the protocol are the minimum times. Gels can be fixed without any problem for longer periods. For example, gels can be fixed overnight, with only one solution change. For ultra rapid fixation, the following procedure can be used:

    a) Fix in 10% acetic acid /40% ethanol for 10 minutes, then rinse for 10 minutes in water *(15)*.

    b) Rinse in 40% ethanol for 2x 10 minutes then in water for 2x 10 minutes. Proceed to step 3.

5. The optimal setup for sensitization is the following: prepare four staining boxes containing respectively the sensitizing thiosulfate solution, water (2 boxes), and the silver nitrate solution. Put the vessel containing the rinsed gels on one side of this series of boxes. Take one gel out of the vessel and dip it in the sensitizing and rinsing solutions (1 minute in each solution). Then transfer to silver nitrate solution. Repeat this procedure for all the gels of the batch. A new gel can be sensitized while the former one is in the first rinse solution, provided that the 1 minute time is kept (use a bench chronometer). When several batches of gels are stained on the same day, it is necessary to prepare several batches of silver solution. However, the sensitizing and rinsing solutions can be kept for at least three batches, and probably more.

6. Gels can be impregnated with silver for at least 30 minutes and at most 2 hours without any change in sensitivity or background.

7. This very short step is intended to remove the liquid film of silver solution brought with the gel.
8. In a standard analytical 2D gel loaded with 50-100µg of protein, the first major spots should begin to appear within 1 minute. Delayed appearance indicates lower than expected sensitivity, but is observed when aldehydes have not been used in the fixing process. In the latter case, sensitivity is restored by a longer development. The developer should be altered if no thiosulfate is present in the gel (e.g. use of ready-made gels). To prevent the rapid appearance of background, add 10 µl of 10% thiosulfate solution per liter of developer. The maximum sensitivity will not be altered by this variation. However, some spots will show a lighter color or give hollow spots.
9. When the gel is dipped in the developer, a brown microprecipitate of silver carbonate should form. This precipitate must be redissolved to prevent deposition and background formation. This is simply achieved by <u>immediate</u> agitation of the box. Do not expect the appearance of the major spots before 3 minutes of development. The spot intensity reaches a plateau after 15-20 minutes of development, and then background appears. Stop development at the beginning of background appearance. This ensures maximal and reproducible sensitivity.
10. The developing solution II is prepared as follows: For one liter of solution, add 0.1 mole of boric acid and 15 0ml of 1 *N* sodium hydroxide to 500 ml water. When the boric acid is dissolved, add 20 g of galactose (or xylose) and finally the sodium thiosulfate. Complete to one liter with water. This solution is not stable and must be prepared the day of use.

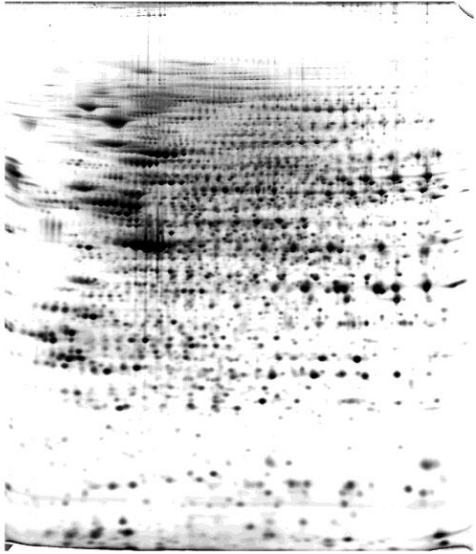 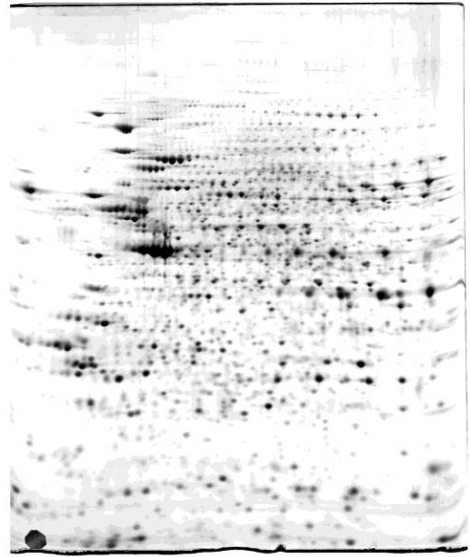

A  B

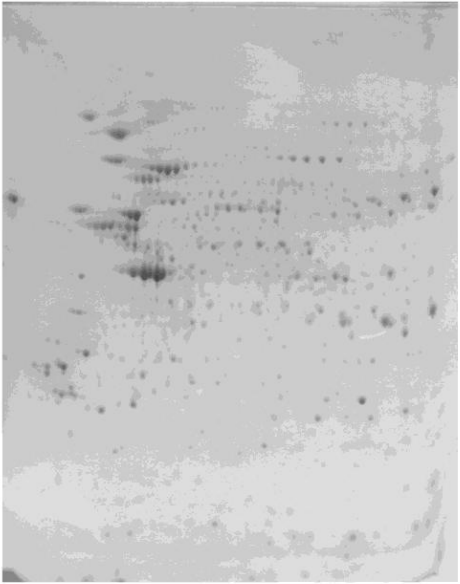

C

Fig. 1: 200µg of proteins extracted from the murine macrophage cell line J774 were loaded on a two dimensional (2D) gel (first IEF dimension: 4-8 linear pH gradient, second dimension: SDS-polyacrylamide gel (10% T)). The protein detection method was the following:
A) silver stain with formaldehyde developer. B) silver stain with xylose-borate developer. C) colloidal Commassie Blue.

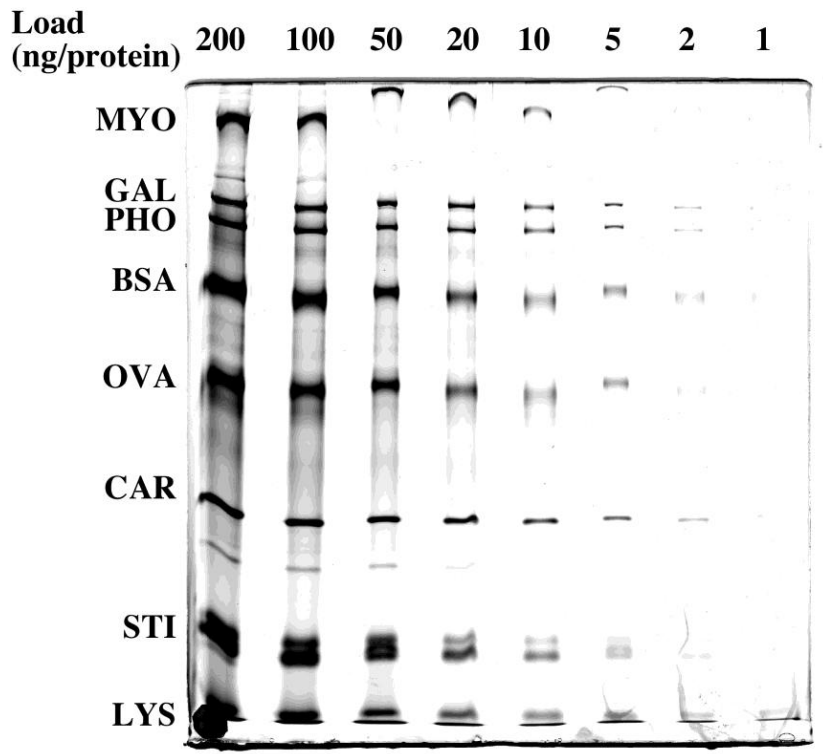
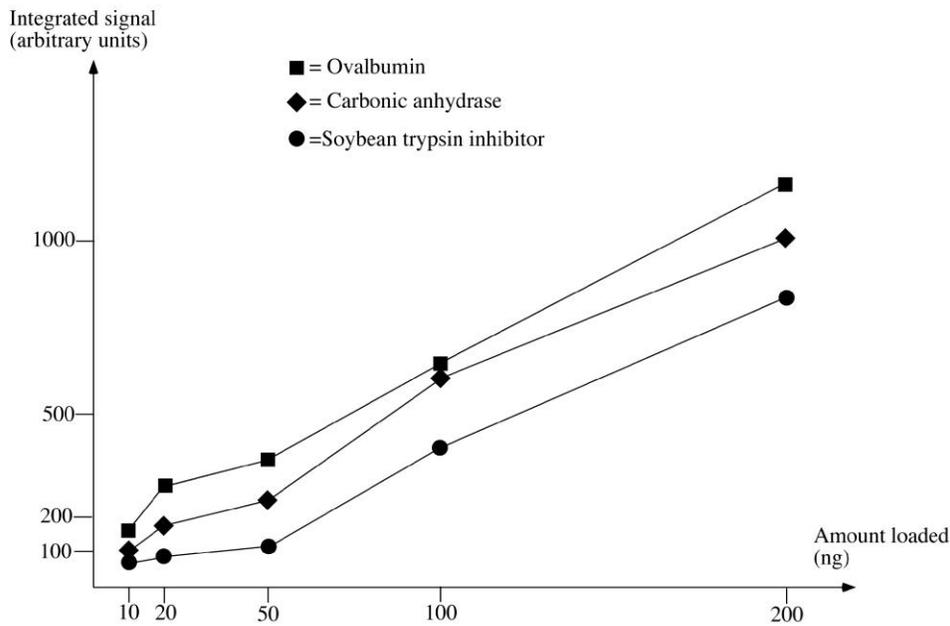

Fig. 2: Molecular mass markers (wide range from BioRad) were serially diluted, loaded on a SDS-polyacrylamide gel (10% T) and silver stained by protocol 1. The resulting image is shown on the top panel. Quantification of three protein bands was then achieved by the ImageJ software (http://rsbweb.nih.gov/ij/) , and the plotted results are shown on the bottom panel. This shows the biphasic shape of the curve, with a plateau at low loads and the linear portion of the curve at higher loads.

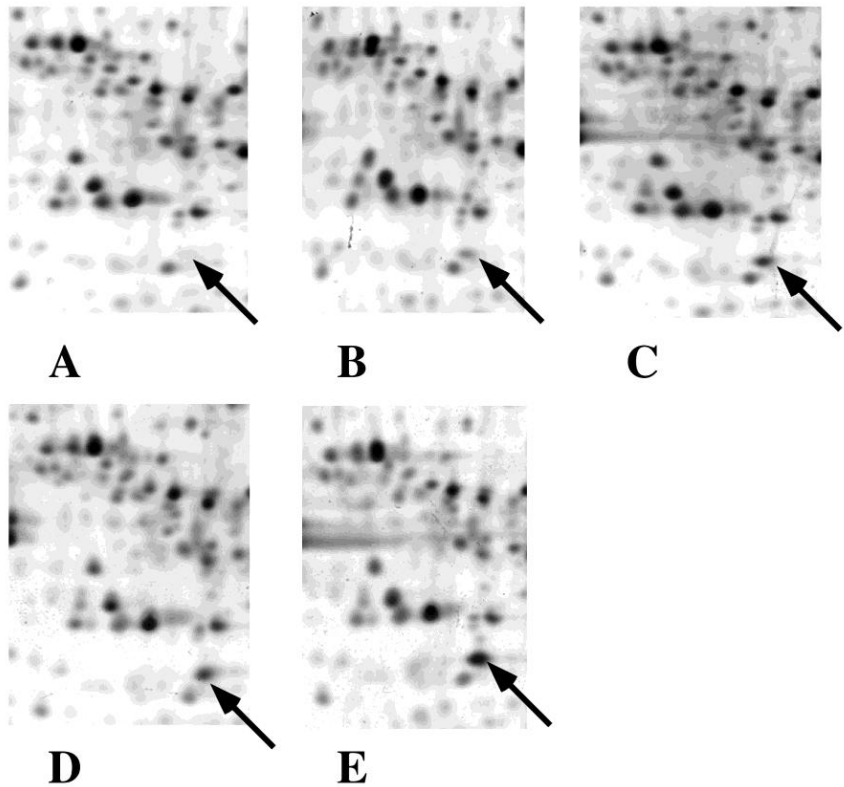

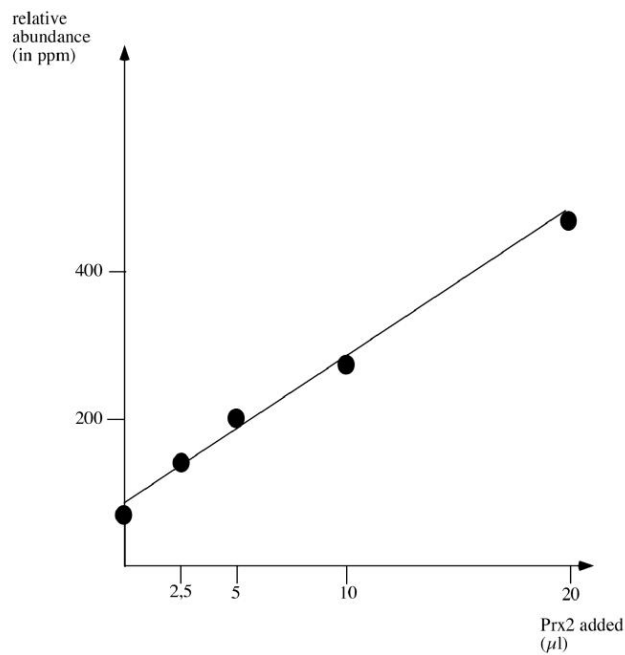

Fig. 3: A total extract from human monocytes was separated by 2D-PAGE (first IEF dimension: 4-8 linear pH gradient, second dimension: SDS-polyacrylamide gel (10% T)), and silver stained (protocol 1). Prior to separation, a variable amount of a semi-purified preparation of peroxiredoxin 2 (prx2) was added to the monocyte extract. The prx2 spot is shown by an arrow in the gel excerpts shown on the figure. A: no prx2 added. B: 2.5 µl prx2 added. C: 5 µl prx2 added. D: 10 µl prx2 added. E: 20 µl prx2 added. The gels were then analyzed by the Delta 2D soltware (Decodon, Greifswald, Germany) and the amount of prx2 determined (in ppm of the total spots intensities). The results are shown in the bottom part of the figure.

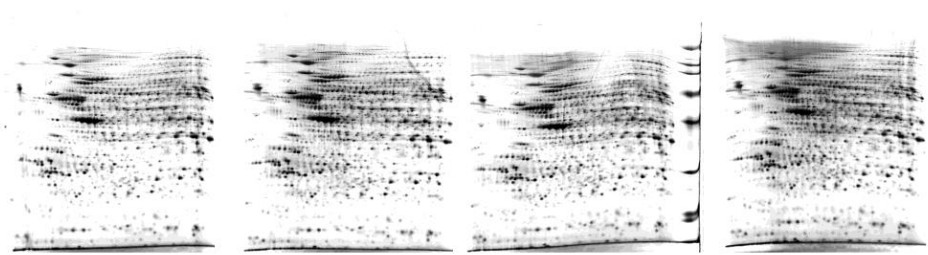

Silver stain protocol 1

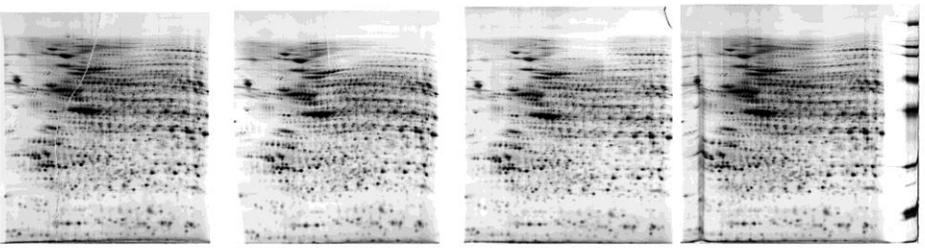

Silver stain protocol 2

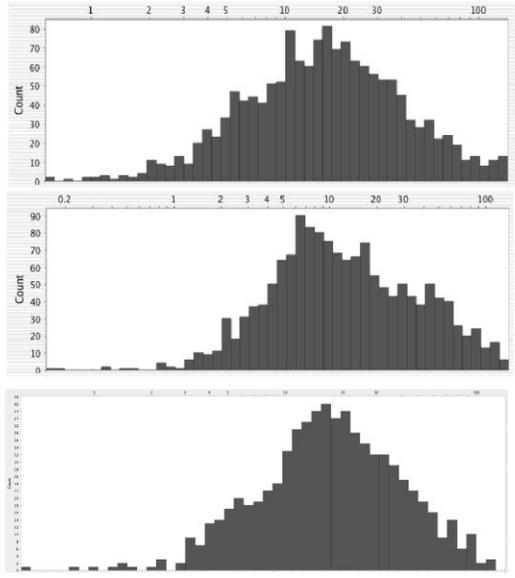

rsd silver protocol 1 median: 20%

rsd silver protocol 2 median: 15%

rsd colloidal Coomassie Blue median: 20%

Fig. 4: A total cell extract from J774 cells was separated in quadruplicate by 2D-PAGE (first IEF dimension: 4-8 linear pH gradient, second dimension: SDS-polyacrylamide gel (10% T)). One four-gel series was stained with silver and formaldehyde developer (protocol 1) and is shown on the top row. The second series was stained with silver and aldose developer (protocol 2) and is shown on the second row. The gels were then analyzed with the Delta 2D software (Decodon) and the relative standard deviation (rsd) was calculated for each spot and is then plotted as a distribution graph (bottom part of the figure). The median of the rsd was calculated, and gives a measure of the variability of the process. This rsd plot is also shown for Coomassie Blue-stained gels for comparison.